\documentclass[conference]{IEEEtran}
\IEEEoverridecommandlockouts
\usepackage{cite}
\usepackage{amsmath,amssymb,amsfonts}
\usepackage{textcomp}
\usepackage{xcolor}
\usepackage{svg}
\usepackage{amsmath}
\def\BibTeX{{\rm B\kern-.05em{\sc i\kern-.025em b}\kern-.08em
    T\kern-.1667em\lower.7ex\hbox{E}\kern-.125emX}}
\usepackage[font=small,labelfont=bf]{caption}
\usepackage{subfigure} 
\usepackage[numbers,sort&compress]{natbib}

\usepackage{hyperref}
\usepackage{float}
\usepackage{stfloats}
\usepackage{graphicx}
\usepackage[linesnumbered,ruled,vlined]{algorithm2e}
\usepackage{xcolor}

\SetAlFnt{\normalsize}
\usepackage{algorithmic}
\algsetup{linenosize=\normalsize}

\SetCommentSty{mycommfont}
\usepackage{booktabs}

\begin{document}

\title{AutoTSMM: An Auto-tuning Framework for Building High-Performance Tall-and-Skinny Matrix-Matrix Multiplication on CPUs
}

\author{\IEEEauthorblockN{
Chendi Li\IEEEauthorrefmark{1}\IEEEauthorrefmark{2},
Haipeng Jia\IEEEauthorrefmark{1}\IEEEauthorrefmark{4}\thanks{\IEEEauthorrefmark{4}Haipeng Jia is corresponding author.},
Hang Cao\IEEEauthorrefmark{1}\IEEEauthorrefmark{2},
Jianyu Yao\IEEEauthorrefmark{1}\IEEEauthorrefmark{2},
Boqian Shi\IEEEauthorrefmark{3},\\
Chunyang Xiang\IEEEauthorrefmark{1}, 
Jinbo Sun\IEEEauthorrefmark{1}\IEEEauthorrefmark{2},
Pengqi Lu\IEEEauthorrefmark{1}\IEEEauthorrefmark{2},
Yunquan Zhang\IEEEauthorrefmark{1}}

\IEEEauthorblockA{\IEEEauthorrefmark{1}Institute of Computing Technology, Chinese Academy of Sciences}
\IEEEauthorblockA{\IEEEauthorrefmark{2}University of Chinese Academy of Sciences}
\IEEEauthorblockA{\IEEEauthorrefmark{3}Indiana University Bloomington}}

\maketitle

\begin{abstract}

In recent years, general matrix-matrix multiplication with irregular-shaped input matrices has been widely used in many applications like deep learning and has drawn more and more attention. However, conventional implementations are not suited for irregular-shaped matrix-matrix multiplications, and few works focus on optimizing tall-and-skinny matrix-matrix multiplication on CPUs. This paper proposes an auto-tuning framework, AutoTSMM, to build high-performance tall-and-skinny matrix-matrix multiplication. AutoTSMM selects the optimal inner kernels in the install-time stage and generates an execution plan for the pre-pack tall-and-skinny matrix-matrix multiplication in the runtime stage. Experiments demonstrate that AutoTSMM achieves competitive performance comparing to state-of-the-art tall-and-skinny matrix-matrix multiplication. And, it outperforms all conventional matrix-matrix multiplication implementations.

\end{abstract}


\section{Introduction}
General matrix-matrix multiplication(GEMM) is the heart of the Basic Linear Algebra Subprograms(BLAS). It is widely used in scientific computing, deep learning, statistics, and many other domains. Many vendors and communities already optimized GEMM on various platforms. There are many open-source implementations including ATLAS\cite{atlas}, OpenBLAS\cite{openblas}, BLIS\cite{blis} and Eigen\cite{eigen}. Meanwhile, some vendor libraries like Intel MKL(i.e. OneAPI)\cite{mkl}, ARMPL\cite{armpl}, AOCL\cite{aocl} and ROCM\cite{rocm} are provided on the specific platform. The formula of GEMM as follows, where A, B and C are the input matrices with the size of \(m\times k\), \(k\times n\), and \(m\times n\), respectively, \(\alpha\) and \(\beta\) are scalar numbers, C is the output matrix,

\begin{equation}
C = \alpha AB  + \beta C
\end{equation}

%For example, Intel MKL optimized for small-scale input on CPUs\cite{LIBXSMM1, LIBXSMM2}. Also, Li proposed a batch GEMM framework\cite{PPoPP19} to speed up small-scale input on GPUs. 
Conventional GEMM implementations are targeting regular-shaped matrices(i.e., m and n are relatively large while k may be moderate). However, the input sizes vary in different applications, and the input sizes are usually irregular-shaped. So, optimization of irregular-shaped GEMM has also drawn a lot of attention.

Tall-and-skinny matrix-matrix multiplication(TSMM) is extensively used in applications like deep learning\cite{facebook, Georganas}.  TSMM means one of the input matrices(A or B) is a tall-and-skinny matrix(one dimension is significantly smaller than another one). Many works have been done on TSMM. For example, Intel MKL provides highly optimized TSMM on the X86 platform\cite{post}, Facebook already optimized TSMM on X86 processors in their data centers\cite{facebook}, TSM2\cite{tsm2} and TSM2X\cite{tsm2x} discussed the TSMM optimization on NVIDIA GPUs. However, TSMM has not been fully discussed, and few works have been done to optimize TSMM on different CPUs. Intel MKL provides high-performance TSMM on the X86 platform, but no TSMM implementation is available on other modern CPU architectures. 

Matrix-matrix multiplications on CPUs are divided into two operations: packing operation and computing operation. The packing operation performs a tiled algorithm that makes memory access continuous during the computation, and the performance can benefit from cache locality. The overhead of packing operation is negligible when computing regular-shaped GEMM because it can be amortized in the computing operation. But conventional GEMM implementations can only achieve sub-optimal performance when computing irregular-shaped matrix-matrix multiplications like TSMM, because the packing operation cannot fully be amortized. In addition, matrix data reuse is needed in deep learning, but conventional GEMM implementations cannot reuse matrices because the packing operation and computing operation are coupled. As a result, a pre-pack TSMM is necessary, and the AutoTSMM packs the input matrices before execution, it makes data available for reuse.

In this paper, we analyze the performance issues with conventional GEMM implementations on computing TSMM. And we propose a novel tiled algorithm and develop a portable auto-tuning framework, AutoTSMM, for building high-Performance TSMM on mainstream CPUs. AutoTSMM selects the optimal inner kernels in the install-time stage and generates an execution plan in the runtime stage. AutoTSMM makes high-Performance TSMM available on mainstream CPUs like X86 platforms and ARMv8 platforms. In addition, we focus on the optimization of inner kernels on ARMv8 platforms to achieve high performance. Experiments show that our work is comparable with state-of-the-art TSMM implementation and get an average speedup from 2.3x to 21.7x comparing to conventional GEMM implementations.

The main contributions of this paper are summarized as follows:
\begin{itemize}
\item We analyze the shortcomings of the conventional GEMM implementations on computing TSMM and propose a novel tiled strategy to implement high-performance TSMM on CPUs.
\item We design an auto-tuning framework, AutoTSMM, for building high-Performance TSMM on all mainstream CPUs. And the performance is competitive with state-of-the-art TSMM implementation from Intel MKL and outperforms all conventional GEMM implementations on X86 and ARMv8 platforms.
\item We analyze the poor performance of TSMM on the ARMv8 platform and implement high-performance assembly inner kernels on the ARMv8 platform.
\end{itemize}

\begin{figure*}
\centering
\includegraphics[width=0.98\textwidth]{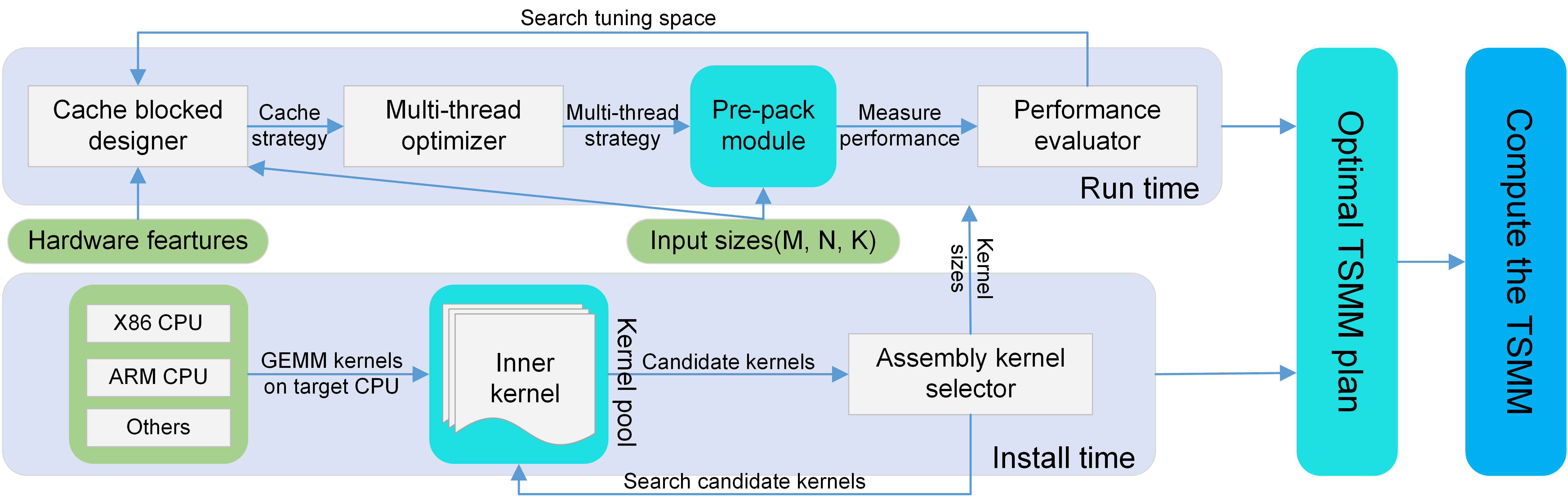}
\caption{Overview of AutoTSMM}
\label{fig:auto}
\end{figure*}

We introduce the background of our works in section 2. AutoTSMM is introduced in section 3. In section 4, we describe the implementation and optimization of the AutoTSMM on CPUs. We evaluate the performance in section 5 and conclude this paper in section 6.

\section{BACKGROUND}
\subsection{Conventional GEMM Implementations}
GEMM is fully optimized in the last decades\cite{ipdps, auto512, auto-KNL, goto, old}, but most of the implementations target regular-shaped matrices, e.g., large and square matrices. The tiled matrix multiplication algorithm\cite{tiled} packs a small block of the whole A and B matrix and then computes the two small blocks of A and B by the highly optimized small-scale GEMM called inner kernels. These inner kernels are usually written in assembly code. The conventional GEMM implementations are effective when the input matrices are both regular-shaped. Usually, conventional GEMM implementations will get a near-peak performance above 90\% of the theoretical peak performance with regular-shaped input.

\subsection{Neural Networks, Convolution and Im2col}
Neural networks(NNs)\cite{extra1} are the foundations of deep learning. In deep learning frameworks like TensorFlow and OneDNN, most running time is spending on convolution layers\cite{jia, why}. So it is important to optimize convolution and many works have been done.

For example, the image to column (Im2col)\cite{im2col} is one of the most popular algorithms for optimizing convolution. Im2col transforms the convolution to GEMM to utilize the fully optimized BLAS libraries. Irregular-shaped GEMM like TSMM is common in real-world deep learning applications, because the convolutional kernels are usually small-scale, and the input images are large-scale, and both of them need to be reused multiple times. Unfortunately, the conventional GEMM implementations are only optimized for regular-shaped GEMM and do not support data reuse.

\subsection{The Pre-Pack TSMM}
TSMM is one of the most important irregular-shaped matrix-matrix multiplications. To our knowledge, A is a large square matrix, and B is a tall-and-skinny matrix that can cover most needs of deep learning applications\cite{tsm2, facebook}. For example, matrix A is the size of \(20480 \times 20480\), and matrix B is the size of \(20480 \times 16\). Conventional GEMM implementations cannot reuse data, but data reuse is a key factor to optimize TSMM in real-world deep learning applications. The pre-pack TSMM reduces the packing overhead in conventional GEMM implementations, and fulfills the need for data reuse by pre-packing the input matrices to a permanent memory address.

Our works focus on optimizing the pre-pack TSMM on CPUs, and it can benefit numerical libraries like NumPy\cite{numpy} for speeding up convolution layers in deep learning frameworks.

\section{AUTO-TUNING FRAMEWORK}

We introduce the auto-tuning framework of TSMM, AutoTSMM, in this section. AutoTSMM is divided into the install-time stage and the runtime stage. It is responsible for selecting the best inner kernels in the install-time stage and generating the execution plan in the runtime stage. The workflow of the AutoTSMM shows in Fig.\ref{fig:auto}.

%It chooses the best plan according to the runtime performance evaluator and applies the execution plan to compute the input TSMM. 

In the install-time stage, the assembly kernel selector selects the optimal kernel based on the experiences. In the runtime stage, the AutoTSMM designs the tiled algorithm based on the cache sizes, inner kernels and the number of threads. Finally, the AutoTSMM selects an execution plan on target CPUs to compute TSMM.

AutoTSMM assembled into the cache blocked designer, multi-thread optimizer, pre-pack module and performance evaluator in the runtime stage.

The cache blocked designer receives hardware features like the cache hierarchy and the cache sizes. It generates the cache blocked strategy based on the hardware features and the inner kernel sizes. We design a predictive model based on our optimization experiences, it searches the tuning space and creates the cache blocked strategy.

The multi-thread optimizer receives the cache blocked strategy and generates the multi-thread strategy. The strategies are applied to the pre-pack module and measured by the performance evaluator. Note that, the number of threads on n-dimension is one of the most important factors for optimizing multi-threaded TSMM, because the n-dimension is much smaller than other dimensions, and assigns too many threads on n-dimension is not a good idea. So, it should be careful to deal with the multi-threading on n-dimension.

The pre-pack module packs input matrices before computing the TSMM. AutoTSMM applies the cache blocked strategy and the multi-thread strategy to the pre-pack module. And the performance is measured by the performance evaluator.

The performance evaluator is responsible for measuring the performance of the pre-pack module and generating the execution plan. It receives the cache blocked strategy and the multi-thread strategy. After measuring the performance of the strategies, the AutoTSMM selects an execution plan to compute the TSMM.

%works in the runtime stage so it may take some time before computing TSMM, but it
Since a bunch of data reuse is needed in real-world applications like deep learning, AutoTSMM certainly benefits the performance of pre-pack TSMM, because the execution plan will be repeatedly executed and the overhead of AutoTSMM will be negligible.

\section{IMPLEMENTATION and OPTIMIZATION}
We design a tiled algorithm\cite{tiled, ipdps} for the pre-pack TSMM and focus on the optimization of the AutoTSMM in this section. Algorithm.\ref{algo:prepack} shows the workflow of the pre-pack TSMM. The AutoTSMM pre-packs the input matrices into a continuous memory, and then it performs the computing operation to compute the TSMM.

\LinesNumberedHidden
\begin{algorithm}
\SetKwBlock{Parallel}{parallel}{end}
\SetKwBlock{DoParallel}{do in parallel}{end}

\SetKwInput{KwInput}{Input}                % Set the Input
\SetKwInput{KwOutput}{Output}              % set the Output
\DontPrintSemicolon
  
  \KwInput{matrices :$A$, $B$ and $C$, scalar number: $M$, $N$, $K$, $\alpha$ and $\beta$, temporary memory address: $DESTA$ and $DESTB$}
  \KwOutput{$C = \alpha  A \times B + \beta C$}

% Set Function Names
  \SetKwFunction{FMain}{TSMM}
  \SetKwFunction{FpackA}{PACKA}
  \SetKwFunction{FpackB}{PACKB}
  \SetKwFunction{Fcompute}{COMPUTE}
 
% Write Function with word ``Function''
  \SetKwProg{Fn}{Function}{:}{}
  \Fn{\FpackA{$\alpha, SRCA, DESTA$}}{
    $SRCA=\alpha SRCA$;\;
    \For{$jc=0$ ; $jc<K$ ; $jc+=k_c$ }{
  	    \For{$ic=0$ ; $ic<M$ ; $ic+=m_c$}{
        	\For{$it=0$ ; $it<m_c$ ; $it+=m_t$ }{
                Pack a block of $\alpha SRCA$ to $DESTA$;\;
            }
        }
    }
  }
  \SetKwProg{Fn}{Function}{:}{}
  \Fn{\FpackB{$\alpha, SRCB, DESTB$}}{
    $SRCB=\alpha SRCB$;\;
    \For{$jc=0$ ; $jc<K$ ; $jc+=k_c$ }{
  	    \For{$ic=0$ ; $ic<N$ ; $ic+=n_c$}{
            Pack a block of $\alpha SRCB$ to $DESTB$;\;
        }
    }
  }
  \SetKwProg{Fn}{Function}{:}{}
  \Fn{\Fcompute{$\beta, C, DESTA, DESTB$}}{
  \DoParallel{ 
    $C=\beta C$;\;
    \For{$jc=0$ ; $jc<K$ ; $jc+=k_c$ }{
        load a block $\alpha A_c$ from $DESTA$\;
  	    \For{$ic=0$ ; $ic<M$ ; $ic+=m_c$}{
        	\For{$it=0$ ; $it<m_c$ ; $it+=m_t$ }{
  	            \For{$kc=0$ ; $kc<N$ ; $kc+=n_c$}{
                    load a block $B_c$ from $DESTB$;\;
            	    $C_c = \alpha  A_c\times B_c+\beta C_c$;\;
            	}
            }
        }
    }
  }
  }
  \SetKwProg{Fn}{Function}{:}{\KwRet}
  \tcp{main function for computing pre-pack TSMM} 
  \Fn{\FMain{$\alpha, \beta, A, B, DESTA, DESTB$}}{
    \tcp{pack $\alpha A$ to memory address $DESTA$} 
        PACKA($\alpha, A, DESTA$);\;
    \tcp{pack $B$ to memory address $DESTB$} 
        PACKB($1, B, DESTB$);\;
    \tcp{compute $C = \alpha  A \times B + \beta C$}
        COMPUTE($\beta, C, DESTA, DESTB$);\;
        \KwRet $C$;\;
  }
\caption{Workflow of the pre-pack TSMM}
\label{algo:prepack}
\end{algorithm}

\subsection{The Runtime Tiled Algorithm}

\begin{figure*}[htbp]
\centerline{\includegraphics[width=0.98\textwidth]{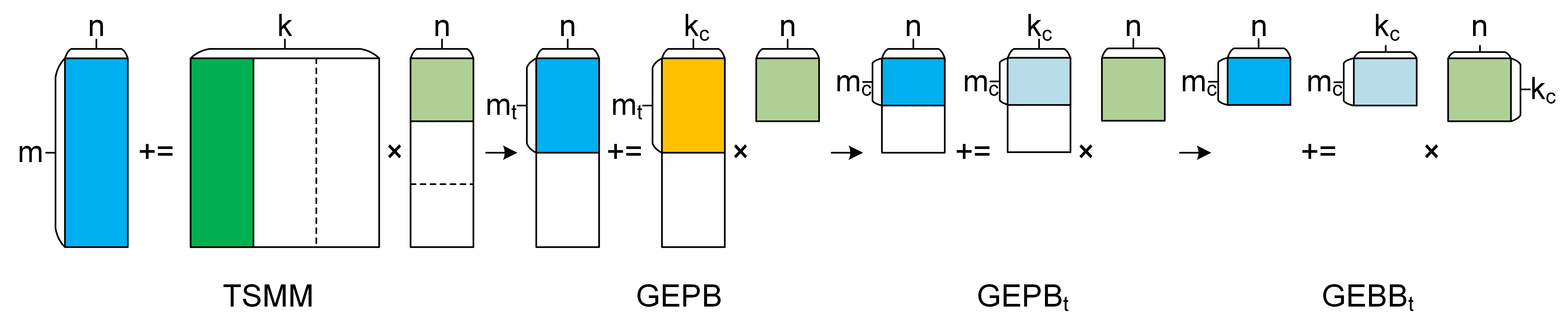}}
\caption{Tiled Algorithm for the Tall-and-Skinny Matrix-Matrix Multiplication. TSMM is transformed to GEPB(panel-block multiplication), where \(m_t\) is block height assigned for one thread, \(k_c\) is the block width suit for L2 cache size. Since n is usually from single digits to hundreds of digits and is significantly smaller than m and k, the n-dimensional tiling algorithm will not be executed when \(n \le n_c\). GEPB is transformed to GEPB\(_t\)(panel-block-multiplication by threads), where \(m_c\) is block height suit for L2 cache size, finally GEBB\(_t\)(block-block-multiplication by threads) is computed as a unit by inner kernels.}
\label{fig:tiled}
\end{figure*}

Since the most difference between regular-shaped GEMM and the TSMM is the length of n-dimension, the tiled algorithm of the TSMM mainly focuses on optimizing the cache blocked sizes and the multi-threading on n-dimension. We design a tiled algorithm for TSMM and the details are illustrated in Fig.\ref{fig:tiled}.

\subsubsection{Cache blocked designer}
GEMM is a cache-oblivious algorithm on CPUs, thus, cache blocked sizes are one of the key factors for optimizing GEMM on modern CPUs. The tiled blocks are selected to fit in L1 cache and L2 cache. For example, assuming we have a \(m_c\times k_c\) sub-matrix A and a \( k_c\times n_c \) sub-matrix B, the best performance of TSMM is achieved most likely when \(m_c\times k_c\) is about half the size of L2 cache and \( k_c\times n_c \) is equal to the size of L1 cache. Thus, the block sizes are limited by Eq.\ref{equation:cache1} and Eq.\ref{equation:cache2} follows. Note that, the \(n_c\), \(m_c\) and \(k_c\) are the block sizes that suit for cache sizes, and \(n_r\), \(m_r\) and \(k_r\) are the block sizes that suit for register sizes.

We design a runtime predictive model to auto-tune TSMM based on the user input sizes. The predictive model based on Eq.\ref{equation:cache1} and Eq.\ref{equation:cache2}, it is responsible for generating the parameters of the tiled algorithm. The search patterns depend on the selected inner kernel and the input. For example, as for \(8\times 4\) kernel, one of the search pattern searches the optimal parameters \(m_{optimal}\)  in \([m_c– 8x,m_c]\) and \(k_{optimal}\)  in \([k_c-4x, k_c]\) according to Eq.\ref{equation:cache1} and Eq.\ref{equation:cache2}. And, another search pattern set the optimal parameters \(m_{optimal}\)  and \(k_{optimal}\) equal to the largest power of two restricted by Eq.\ref{equation:cache1} and Eq.\ref{equation:cache2}. Finally, the performance evaluator chooses the execution plan created by the cache blocked designer and the multi-thread optimizer.

\begin{equation}
 k_c\times n_c\leq \frac{L1cache}{FPsize}\label{equation:cache1}
\end{equation}

\begin{equation}
m_c\times k_c   \leq \frac{L2cache}{2\times FPsize} \label{equation:cache2}
\end{equation}

\subsubsection{Multi-thread optimizer}
The conventional GEMM implementations usually assume the length of n is equal to m, and it will assign the number of threads according to a fixed parameter, thus, conventional GEMM implementations may over-assign threads on n-dimension when computing TSMM. And the over-assignment causes too much thread synchronization overhead and significantly affects multi-threaded performance on computing TSMM.

Meanwhile, the TSMM tiled algorithm in Fig.\ref{fig:tiled} has not divided into slices size into \(m_c\times n_t\) and are computed by GEBB\(_t\) instead of computed by GEBP\(_t\) (block-panel-multiplication by threads) in conventional GEMM implementations. When computing TSMM, divide the n-dimension into slices like conventional GEMM implementations is not a good idea. This is because \(n\) is too small and the blocked sub-panel of matrix B hold by single-thread is much smaller than L1 cache size, so the L1 data cache cannot be fully utilized. AutoTSMM will not divide n-dimension when \(n\le n_c\), and ensure every CPU core holds a block of matrix B in GEBB\(_t\) and keep it in the private L1 data cache. Compare to conventional GEMM implementations, this is a good approach to fully utilize the L1 data caches when computing TSMM.
% and reduce the overhead caused by threads over-assignment 

In addition, the packing operation in conventional GEMM implementations allocates temporary space to store continuous panels or blocks of matrix. To prevent the temporary space from being overwritten by the multi-threading packing operation, each thread stall until all threads finish the GEBPs, then threads continue to iterate on the K dimension. As for the pre-pack TSMM, each thread can directly compute the tiled matrix-matrix multiplication without thread synchronization, because every block has been packed into a new memory address. The pre-pack TSMM is a trade-off, but it is effective when the data is reused.

\subsubsection{Pre-pack TSMM module}

\begin{figure}[htbp]
\centerline{\includegraphics[width=0.48\textwidth]{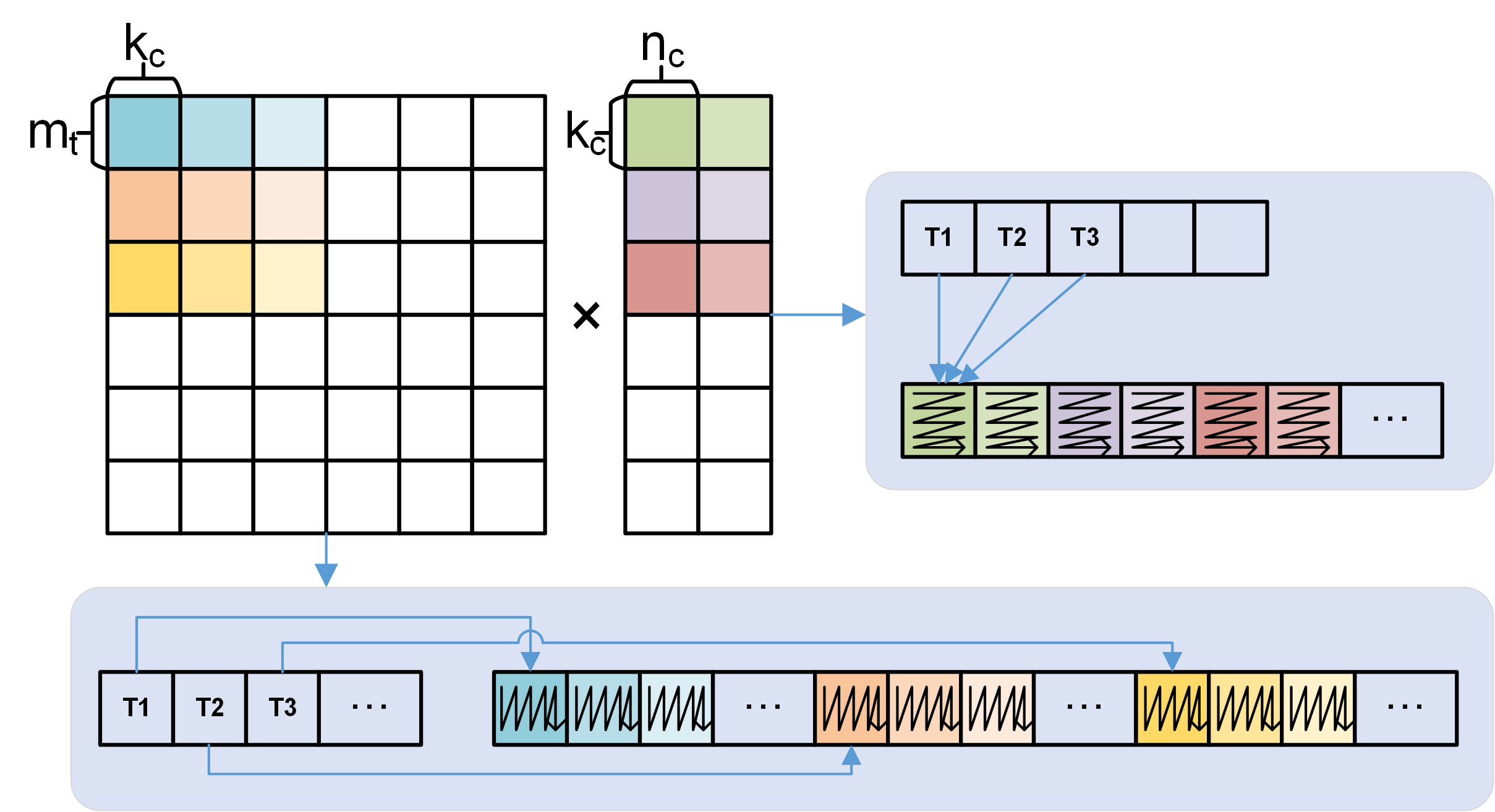}}
\caption{Workload of The Pre-Pack Module}
\label{fig:memory}
\end{figure}

The packing operation is responsible for re-arranging the input matrices to reduce the L1 cache misses. As shown in section 5, the packing time is really large when computing TSMM, so it is important to implement the pre-pack TSMM for data reuse. The workload of the pre-pack module and the memory diagram of the packed matrix are illustrated in Fig.\ref{fig:memory}. We optimize the memory access to minimize the cache misses for each thread. The pre-pack TSMM module copies a block of the input matrix to a new memory address and makes the memory access continuous when computing GEBB$_t$ by inner kernels. As shown in Fig.\ref{fig:memory}, the input matrices are packed into the new continuous memory, and the headers keep the pointers to the address of every block assigned to the threads. Note that, the blocks in Fig.\ref{fig:memory} represents the GEBB\(_t\) in the TSMM tiled algorithm. And, the sizes of the blocks are determined by the cache blocked designer.

\subsubsection{Cache miss analysis for pre-pack TSMM}
Cache complexity\cite{cache} is suitable for analyzing the cache miss rate. To simplify the analysis, we assume all the matrices are stored by column-major, and \(m_c=n_c=k_c=b\), \(m=n=k=n\), the cache size is Z and the cache is divided into cache-lines of size L. And T is the number of threads. Note that, this is not an accurate analysis for TSMM, but it shows how the reduction of cache complexity occurs in pre-pack TSMM. Based on the analysis from \cite{complexity}. Targeting on computing the basic 3-nested loop GEMM, we compute L elements of C will spend \(n^2/L\) times, and it will cause n cache misses each time, thus, the cache complexity of basic 3-nested loop GEMM is:

\begin{equation}
O(n^3/L)
\end{equation}

As for matrix tiling in conventional GEMM implementations(i.e. blocking GEMM), loading three \(b\times b\) blocks in cache cost \(3b^2/L\) for \(n^3/b^3\) times. This leads to \(3n^3/(bL)\) cache misses. And three blocks must fit in cache for best performance: \(3b^2<Z\). So the cache complexity of block GEMM is:

\begin{equation}
O(3\sqrt{3}n^3/(L\sqrt{Z})) \label{equation:old}
\end{equation}

As for pre-pack TSMM, assume we pre-pack matrix A, when the computing operations are called, loading A into the cache will only cause \(b\times b\) cache misses when the first sub-matrix A is loaded into the cache. It will cause \(b^2/L\) cache misses for every thread, and loading blocks from B and C cost \(2b^2/L\) for \(n^3/b^3\) times, and we already get \(3b^2<Z\). So the cache complexity of computing pre-pack TSMM is:

\begin{equation}
O(ZT/(3L)+2\sqrt{3}n^3/(L\sqrt{Z}))\label{equation:new}
\end{equation}

According to Eq.\ref{equation:old} and Eq.\ref{equation:new}, and comparing to conventional GEMM implementations, pre-pack TSMM reduces the cache misses caused by packing operation. And the performance is benefits from the data rearrangement and the data reuse.

\subsection{The Install-time GEMM Kernels Optimization}

\begin{figure}[htbp]
\centerline{\includegraphics[width=0.48\textwidth]{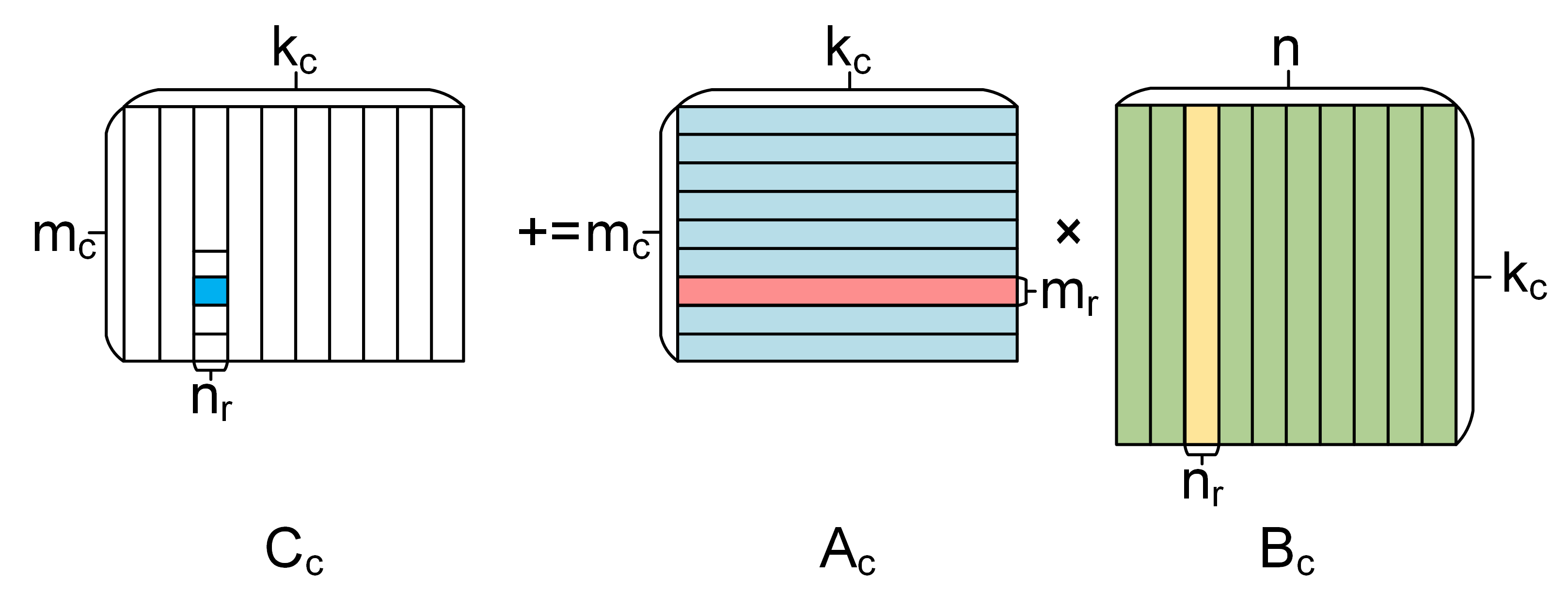}}
\caption{GEBB\(_t\) Computed by Inner Kernels. The inner kernel perform a slice-times-slice matrix-matrix multiplication(\(m_r\) and \(n_r\) are the sizes suit for register blocking).}
\label{fig:GEBB}
\end{figure}

Fig.\ref{fig:GEBB} shows how to compute GEBB\(_t\) by inner kernels. Since the inner kernels on the X86 platforms have been fully optimized, We reuse the inner kernels on X86 platforms to implement the AutoTSMM. However, the inner kernels on ARMv8 platforms have not been fully optimized and the existing ARMv8 \(16\times 4\) and \(8\times 4\) inner kernels cannot achieve high-performance. Meanwhile, the flaws on Kunpeng 920 make the existing inner kernels can only achieve sub-optimal performance. Thus, we focus on the optimization of assembly inner kernels on Kunpeng 920 and try to figure out how to improve the poor performance caused by the instruction issue flaws.

Kunpeng 920 platform has two fused multiply-add (FMA) units, ideally, peak performance is achieved when two FMA instructions and one or more load instructions are issued every cycle. However, Kunpeng 920 can only issue two single-precision FMA instructions per cycle or issue one FMA instruction and one load instruction per cycle. Thus, the peak performance of Kunpeng 920 is limited by the memory access instructions(i.e., load instructions).

As discussed above, the FMA instructions cannot overlap the load instructions on Kunpeng 920, so we have to minimize the percentage of the load instructions when implementing the inner kernels. Kunpeng 920 has 32 SIMD register files, we assume the inner kernel size is \(m_r\times n_r\), we implement \(12\times 8\), \(16\times 4\) and \(8\times 4\) kernels on Kunpeng 920 platform. The \(12\times 8\) inner kernel has the largest FMA instructions ratio 92.3\%, so it is the optimal inner kernel. This rule is not suited for X86 CPUs because the flaws are only found in Kunpeng 920. In addition, the \(12\times 8\) inner kernel can output more elements of matrix C than the \(16\times 4\) inner kernel at a time, which means that its utilization of registers is higher. Now we introduce the details of \(12\times 8\) inner kernel optimization on Kunpeng 920.

\subsubsection{The register blocking}
There are 32 SIMD vector registers on Kunpeng 920 platform, as for \(12\times 8\) inner kernel, we use V0, V1 and V6, V7 to store the matrix A, V2-V5 are used for storing matrix B, V8-V31 are used for storing matrix C. Only V6-V7 are used exclusively, other registers are used for register renaming.

\subsubsection{Reorder instructions and loop unrolling}
The Kunpeng 920 cannot fulfill the pipeline, so we reorder instructions and let the FMA instructions and load instructions are spaced farther apart to avoid pipeline hazards. We assemble the inner kernels by KERNEL\_M1, KERNEL\_M2. This is a ping-pong strategy, which is an effective way to build inner kernels. It means KERNEL\_M1 loads the data for KERNEL\_M2, and KERNEL\_M2 also loads the data for KERNEL\_M1. We use loop-unrolling on k-dimension by repeatedly calling KERNEL\_M1 and KERNEL\_M2 to make the pipeline throughput higher.

\subsubsection{Prefetch from the cache} 
The prefetch instructions are also used for optimization. We insert the prefetch instructions between floating-point instructions, as for \(12\times 8\) kernel, it prefetches 5120 bytes, 448 bytes and 320 bytes for matrix A, B and C, respectively. Prefetching seems likely to be a hand-writing optimization on target machines.

\section{PERFORMANCE EVALUATION}
This section shows the experimental results on X86 and ARMv8 platforms. We compare the performance of AutoTSMM on computing single-precision TSMM (STSMM) and double-precision TSMM (DTSMM) with state-of-the-art implementations. Table 1 shows the experimental environments.

\begin{table}[htbp]
\begin{center}
\label{env}
\caption{Experimental Environments}
\begin{tabular}{ccc}
\toprule
CPU & Kunpeng 920 & Xeon E5-2640 v4\\
\midrule
Arch. & ARMv8.2 & Broadwell-EP\\
Freq. & 2.6GHz& 2.4 GHz\\
SIMD & 128 bits& 256 bits\\
L1D cache & 4 MiB& 320 KiB	\\
L2 cache & 32 MiB& 2.5 MiB\\
L3 cache & 64 MiB& 25 MiB\\
Compiler & GCC7.5& GCC7.5\\
Intel MKL & - & 2021.2.0\\
ARMPL & 21.0 & - \\
BLIS & 0.81& 0.81\\
OpenBLAS & 0.3.13 & 0.3.13\\
\bottomrule
\end{tabular}
\end{center}
\end{table}

\begin{figure*}[htbp]
\centering
\begin{minipage}[b]{.48\textwidth}
\includegraphics[width=\textwidth]{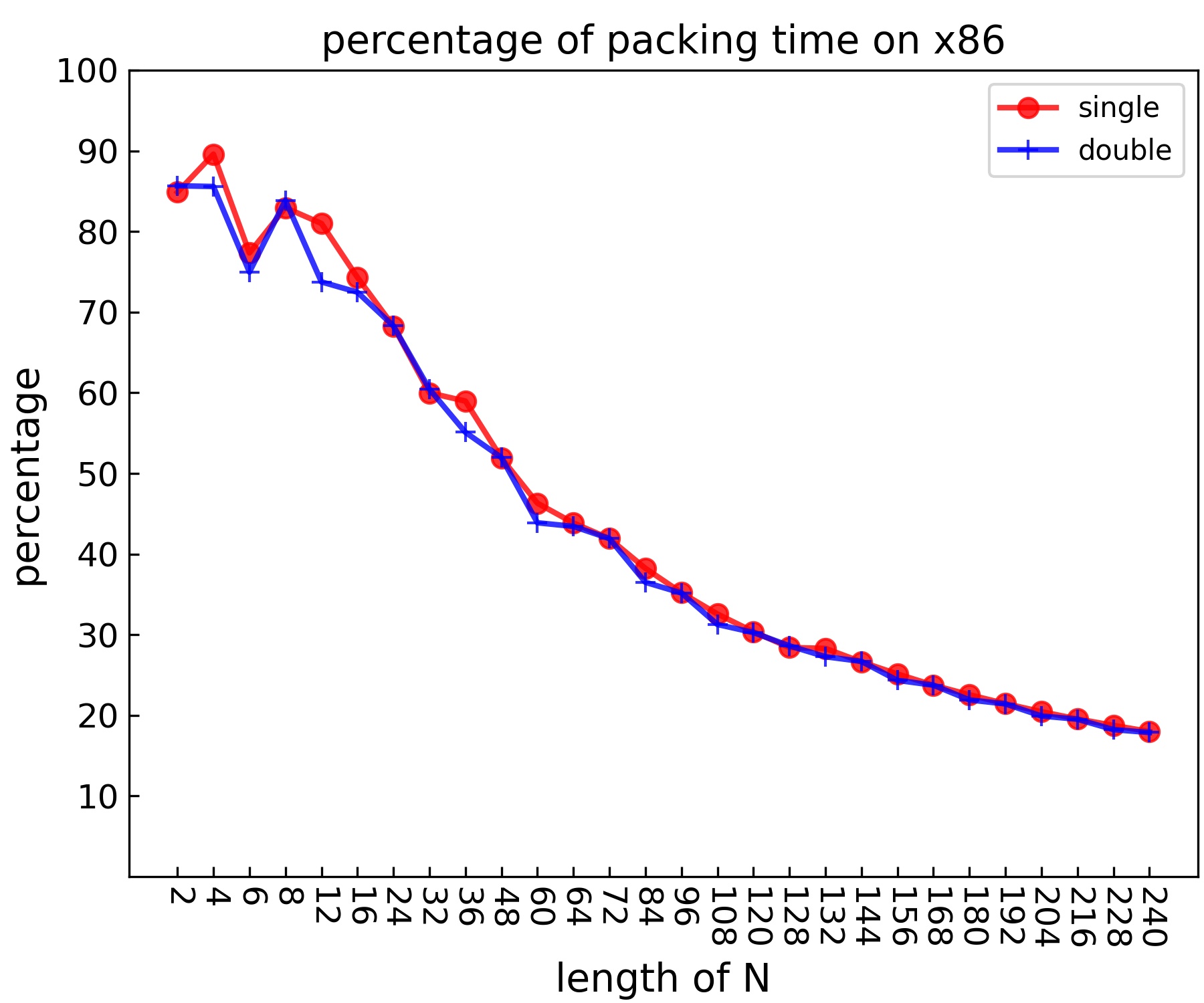}
\end{minipage}\qquad
\begin{minipage}[b]{.48\textwidth}
\includegraphics[width=\textwidth]{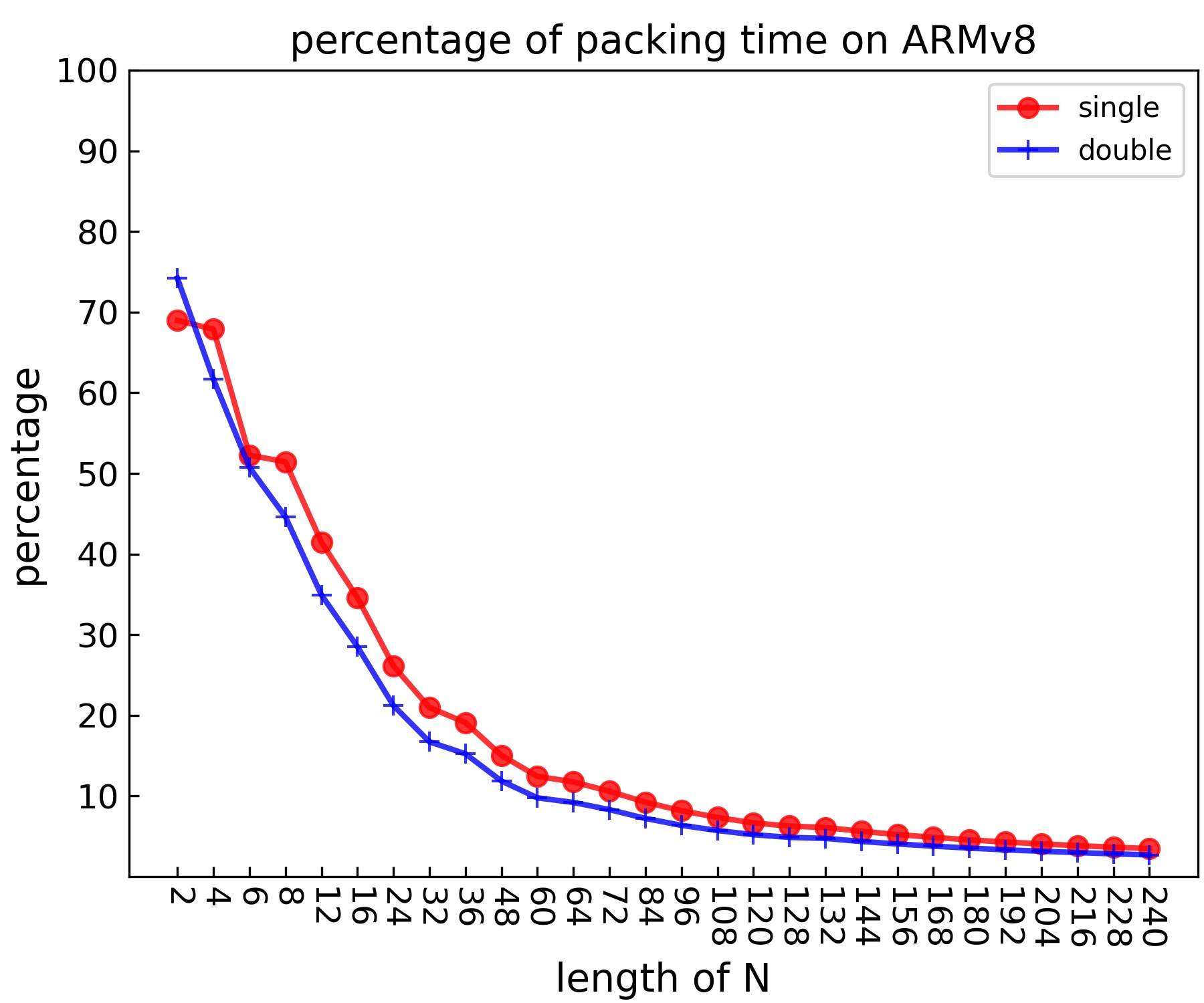}
\end{minipage}
\caption{The Percentage of the Packing Operation Time in Conventional GEMM Implementation on X86 and ARMv8 CPUs}
\label{fig:occupation}

\begin{minipage}[b]{.48\textwidth}
\includegraphics[width=\textwidth]{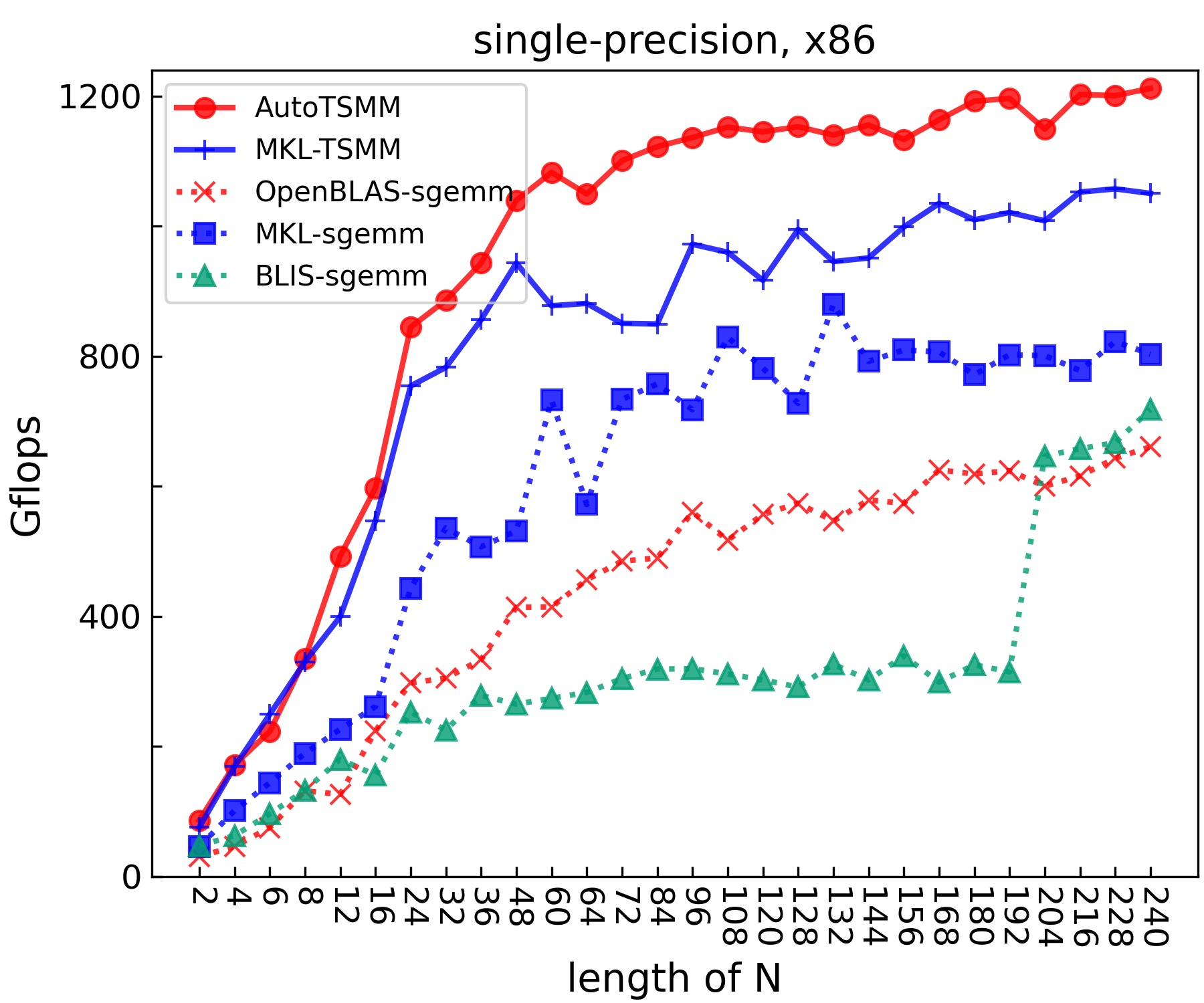}
\end{minipage}\qquad
\begin{minipage}[b]{.48\textwidth}
\includegraphics[width=\textwidth]{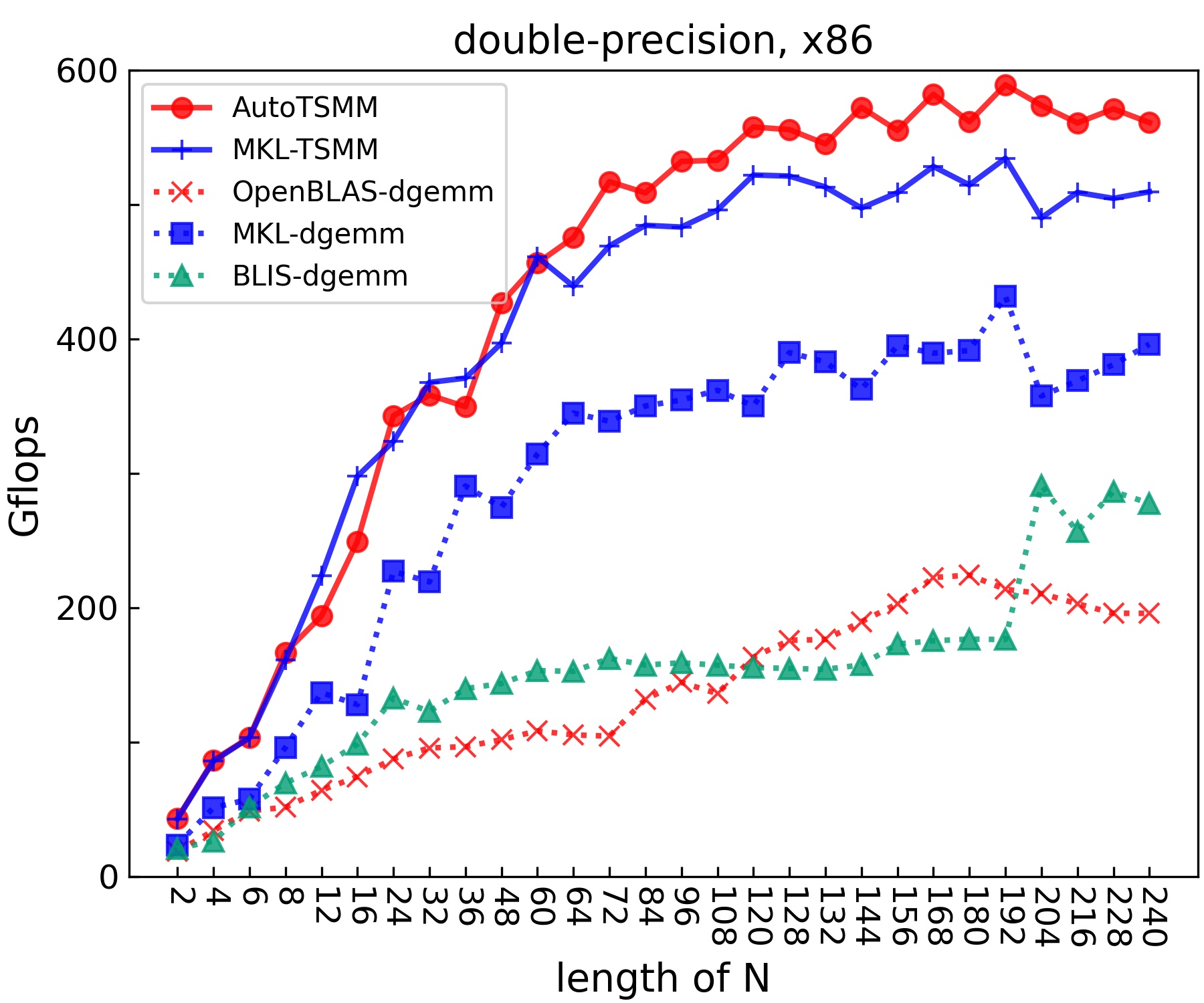}
\end{minipage}
\caption{The TSMM Performances on X86 CPUs}
\label{fig:X86perf}

\begin{minipage}[b]{.48\textwidth}
\includegraphics[width=\textwidth]{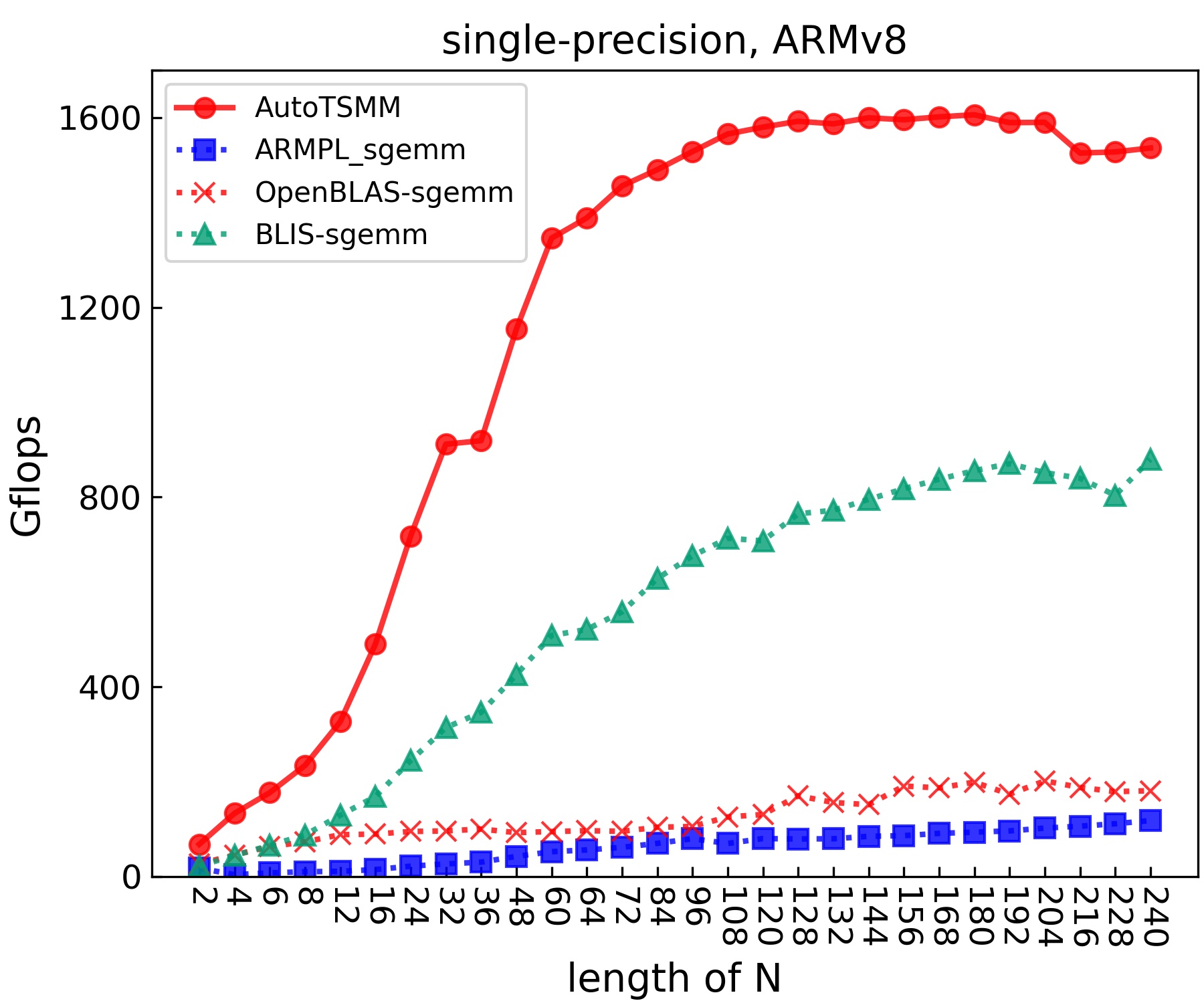}
\end{minipage}\qquad
\begin{minipage}[b]{.48\textwidth}
\includegraphics[width=\textwidth]{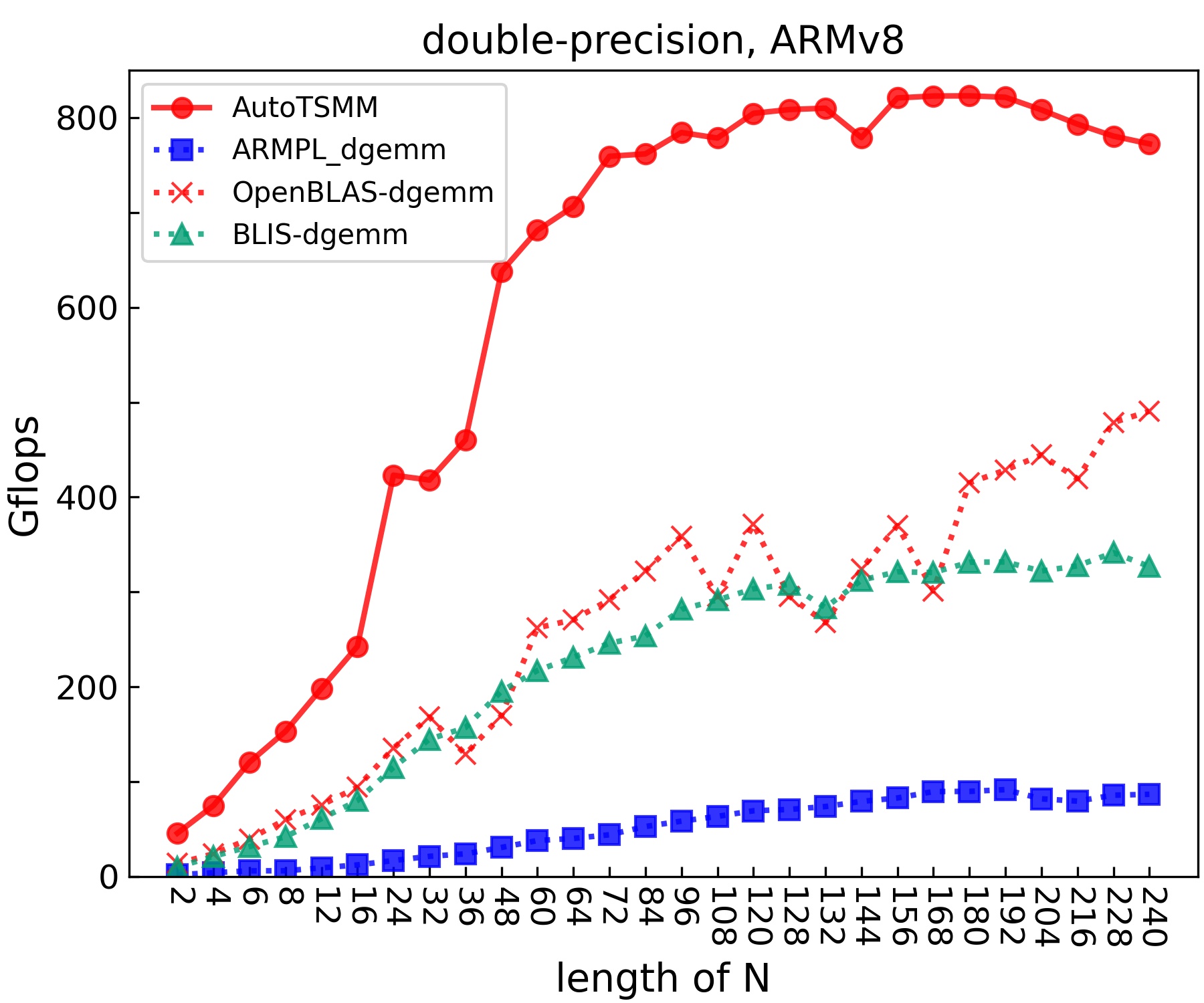}
\end{minipage}
\caption{The TSMM Performances on ARMv8 CPUs}
\label{fig:armperf}
\end{figure*}

We assume that matrix A is a large square matrix, and matrix B is a tall-and-skinny matrix. Note that, we choose these test cases because we believe they can show the challenge in optimizing TSMM. The performance evaluation computes TSMM 200 times, and \(M=K=25600\), referring to previous works like MKL\cite{post} and TSM2\cite{tsm2}, we believe these test cases are reasonable for comparing the performance of the TSMM. The performance of the AutoTSMM framework is labeled AutoTSMM, The performance of Intel MKL TSMM modules is labeled MKL-TSMM. The conventional GEMM implementations' performances are represented by the BLAS library name plus “sgemm” or “dgemm”, where “sgemm” represents single-precision GEMM(SGEMM), and “dgemm” represents double-precision GEMM(DGEMM). We use Eq.\ref{equation:gflops} for measuring the performance. Notice that, the packing time is ignored.

\begin{equation}
GFlops=\frac{2\times M \cdot N \cdot K \times 10^{-9}}{t}\label{equation:gflops}
\end{equation}

\subsection{Targeting on Intel Xeon}
%Figure 5 shows the percentage of time for packing operations on X86 and ARMv8 platforms, The time occupation of the packing operation is the major overhead in TSMM,  It demonstrates the importance of pre-pack and data reuse when computing TSMM.

Fig.\ref{fig:occupation} shows the packing time taken on the X86 and ARMv8 platforms. When n is very small, the percentage of the packing time is very large, reaching 90\% and 75\%, respectively. With the increase of n, the percentage of the packing time gradually decreases. When $N=240$, the percentage of packing time on the X86 platform is about 20\%, and on the ARMv8 platform is 3\%. In a word, when the length of N is small, the packing time is the main overhead, as the length of N increases, the packing time takes up less.

Fig.\ref{fig:X86perf} shows the performance comparison on X86 platforms. Xeon E5-2640 v4 has 10 cores with 20 threads, so it has 1536 GFlops peak performance for single-precision and 768 GFlops for double-precision. The performance of AutoTSMM can achieve 78.9\% of single-precision peak performance along with 76.7\% of the double-precision peak performance at most. AutoTSMM can achieve 115.2\% and 105.1\% of the Intel MKL-TSMM on computing STSMM and DTSMM, respectively. We believe the reason why DTSMM performance is not as good as STSMM is that the Intel MKL implements the inner kernels with a larger loop-unrolling on K-dimension.

And the conventional GEMM implementations cannot compete with AutoTSMM and MKL-TSMM. On the X86 platform, AutoTSMM achieves 1.63x, 2.38x and 3.19x average speedup comparing to MKL-sgemm, OpenBLAS-sgemm and BLIS-sgemm, respectively. And it achieves 1.53x, 3.27x and 2.87x comparing to MKL-dgemm, OpenBLAS-dgemm and BLIS-dgemm, respectively.

In addition, when \(N=240\), the packing time taken is less than 20\% and 3\% on X86 and ARMv8 platforms, respectively, AutoTSMM can still achieve 30\% to 162\% acceleration compared to conventional implementations. Therefore, we can conclude that our tiled algorithm works well.

From the curves in Fig.\ref{fig:occupation} and Fig.\ref{fig:X86perf}, we can conclude:
\begin{enumerate}
\item AutoTSMM and MKL-TSMM reduce the packing time by pre-pack operation and achieve a significant performance improvement by the data reuse. 
\item  Our runtime tiled algorithm is effective, and it outperforms the Intel MKL on computing STSMM and achieves competitive performance on computing DTSMM. 
\item AutoTSMM brings a significant speedup comparing to conventional GEMM implementations.
\end{enumerate}

\subsection{Targeting on Kunpeng}
\begin{figure}[H]
\centerline{\includegraphics[width=0.47\textwidth]{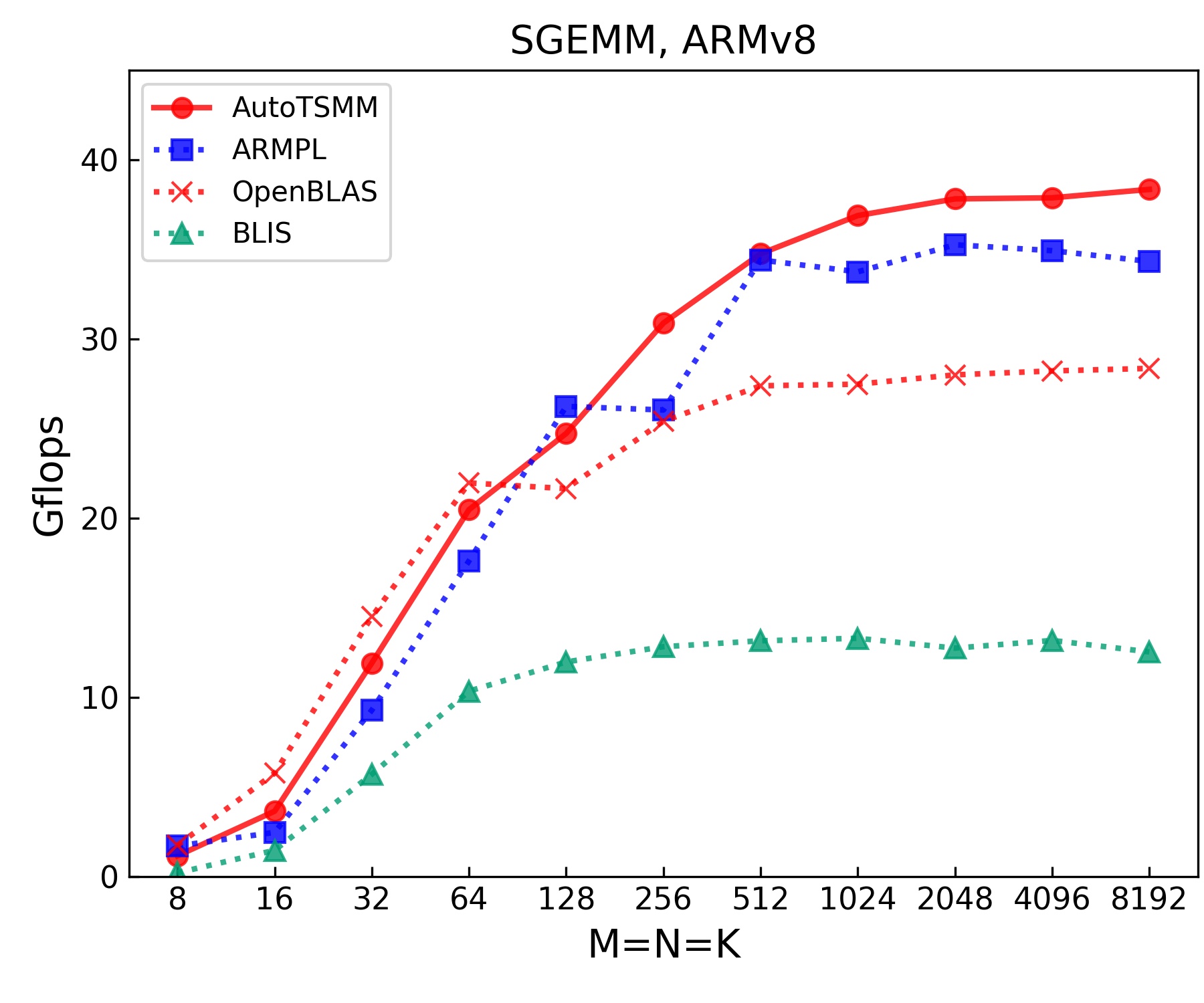}}
\caption{The SGEMM Performances on ARMv8 CPUs}
\label{fig:kernelperf}
\end{figure}
Figure \ref{fig:kernelperf} shows the single-thread SGEMM performance comparison on Kunpeng 920. When the matrix size is small, our \(12\times 8\) inner kernel is not as good as OpenBLAS \(16\times 4\) inner kernel, as the scale increases, our implementation can achieve higher peak performance. The single-core peak performance of Kunpeng 920 is 41.6 GFlops, and our implementations can reach 92.2\% of the peak performance. Our \(12\times 8\) inner kernel has 1.35x, 1.12x and 3.05x peak performance speedup comparing to OpenBLAS, ARMPL and BLIS, respectively. It proves that our optimization of the inner kernels on Kunpeng 920 is effective.

Figure \ref{fig:armperf} shows the performance comparison between AutoTSMM and other implementations on Kunpeng 920. Since there is no open-source or vendor-provided TSMM implementation on the ARMv8 platforms to our knowledge, we compare the AutoTSMM with the conventional GEMM implementations. The performance of AutoTSMM is significantly better than the existing GEMM implementations on the ARMv8 platform.

On the ARMv8 platform, AutoTSMM achieves an average acceleration ratio from 2.3x to 21.7x and 2.9x to 15.1x compared to conventional SGEMM and DGEMM implementations, respectively. We believe that there are two reasons. On the one hand, the existing multi-threaded GEMM implementations are not specifically optimized for the TSMM. On the other hand, our inner kernels are fully optimized to achieve higher performance on Kunpeng 920.

In addition, the existing GEMM implementations do not have a runtime auto-tune process, while AutoTSMM generates the execution plan for better performance. AutoTSMM allows developers to avoid manual tuning and auto-tune for the input sizes, which is one of the most advantages. AutoTSMM is located on the upper layer of the specific architecture, as long as the inner kernels are highly optimized, the high-performance TSMM will surely be achieved on AVX-512 machines and other platforms.

\section{CONCLUSION}
This paper focuses on the implementation and optimization of the auto-tuning framework, AutoTSMM, on X86 and ARMv8 platforms. We can conclude that AutoTSMM is a portable auto-tuning framework to build the high-performance TSMM on mainstream CPUs. When new architecture arrives, the developer does not need to know every detail of the optimization of the TSMM. The only required is the inner kernels on target machines. Thus, the AutoTSMM significantly reduces the effort to apply high-performance TSMM to new platforms. The limitation of AutoTSMM is the number of data reuses. If the number of data reuses is less, then the improvement brought by AutoTSMM will be less obvious.

\section*{Acknowledgment}
We would like to express our gratitude to all reviewer’s constructive comments for helping us polish this article. This work is supported by the National Key Research and Development Program of China under Grant Nos. 2017YFB0202105, the National Natural Science Foundation of China under Grant No. 61972376 and the Natural Science Foundation of Beijing under Grant No. L182053.


\begin{thebibliography}{00}
\bibitem{tsm2} Chen, Jieyang, et al. TSM2: optimizing tall-and-skinny matrix-matrix multiplication on GPUs. Proceedings of the ACM International Conference on Supercomputing. 2019.

\bibitem{tsm2x} Rivera C, Chen J, Xiong N, et al. TSM2X: High-performance tall-and-skinny matrix-matrix multiplication on GPUs[J]. Journal of Parallel and Distributed Computing, 2021, 151: 70-85.

\bibitem{blis}Van Zee F G, Van De Geijn R A. BLIS: A framework for rapidly instantiating BLAS functionality[J]. ACM Transactions on Mathematical Software (TOMS), 2015, 41(3): 1-33.

%\bibitem{LIBXSMM1}Heinecke A, Pabst H, Henry G. Libxsmm: A high performance library for small matrix multiplications[J]. Poster and Extended Abstract Presented at SC, 2015, 15.

%\bibitem{LIBXSMM2}Heinecke A, Henry G, Hutchinson M, et al. LIBXSMM: accelerating small matrix multiplications by runtime code generation[C]//SC'16: Proceedings of the International Conference for High Performance Computing, Networking, Storage and Analysis. IEEE, 2016: 981-991.

%\bibitem{PPoPP19}Li X, Liang Y, Yan S, et al. A coordinated tiling and batching framework for efficient GEMM on GPUs[C]//Proceedings of the 24th Symposium on Principles and Practice of Parallel Programming. 2019: 229-241.

\bibitem{Georganas}Georganas E, Avancha S, Banerjee K, et al. Anatomy of high-performance deep learning convolutions on simd architectures[C]//SC18: International Conference for High Performance Computing, Networking, Storage and Analysis. IEEE, 2018: 830-841.

\bibitem{facebook}Park J, Naumov M, Basu P, et al. Deep learning inference in facebook data centers: Characterization, performance optimizations and hardware implications[J]. arXiv preprint arXiv:1811.09886, 2018.

\bibitem{openblas} Xianyi, Zhang, Wang Qian, and Zhang Yunquan. Model-driven level 3 BLAS performance optimization on Loongson 3A processor. 2012 IEEE 18th international conference on parallel and distributed systems. IEEE, 2012.

\bibitem{mkl}“Accelerate Fast Math with Intel® oneAPI Math Kernel Library.” https://software.intel.com/content/www/us/en/develop/tools/oneapi/comp
onents/onemkl.html

\bibitem{atlas}ATLAS C, Yamamoto S, Shapiro M, et al. The simulation principle and performance of the ATLAS fast calorimeter simulation FastCaloSim[R]. ATL-COM-PHYS-2010-838, 2010.

\bibitem{armpl} “Arm Compiler for Linux | Commercial ArmPL: Get started – Arm Developer.” https://developer.arm.com/tools-and-software/server-and-hpc/compile/arm-compiler-for-linux/resources/get-started/armpl-get-started

\bibitem{aocl} “AMD Optimizing CPU Libraries (AOCL) - AMD.” https://developer.amd.com/amd-aocl/

\bibitem{rocm}“AMD ROCm Open Software Platform | AMD.” https://www.amd.com/en/graphics/servers-solutions-rocm

\bibitem{eigen}“Eigen.” https://eigen.tuxfamily.org/index.php

\bibitem{post} “Unleash The Power Of Big Data Analytics And Machine Learning.” https://www.codeproject.com/Articles/1151600/Unleash-The-Power-Of-Big-Data-Analytics-And-Machin

%\bibitem{tensorflow} Abadi M, Barham P, Chen J, et al. Tensorflow: A system for large-scale machine learning[C]//12th {USENIX} symposium on operating systems design and implementation ({OSDI} 16). 2016: 265-283.

\bibitem{ipdps}Smith T M, Van De Geijn R, Smelyanskiy M, et al. Anatomy of high-performance many-threaded matrix multiplication[C]//2014 IEEE 28th International Parallel and Distributed Processing Symposium. IEEE, 2014: 1049-1059.

\bibitem{auto512}Kim R, Choi J, Lee M. Optimizing parallel GEMM routines using auto-tuning with Intel AVX-512[C]//Proceedings of the International Conference on High Performance Computing in Asia-Pacific Region. 2019: 101-110.

\bibitem{auto-KNL}Lim R, Lee Y, Kim R, et al. Auto-tuning GEMM kernels on the Intel KNL and Intel Skylake-SP processors[J]. The Journal of Supercomputing, 2019, 75(12): 7895-7908.

\bibitem{goto}Goto K, Geijn R A. Anatomy of high-performance matrix multiplication[J]. ACM Transactions on Mathematical Software (TOMS), 2008, 34(3): 1-25.

\bibitem{old}Kågström B, Ling P, Van Loan C. GEMM-based level 3 BLAS: high-performance model implementations and performance evaluation benchmark[J]. ACM Transactions on Mathematical Software (TOMS), 1998, 24(3): 268-302.

\bibitem{tiled}Buttari A, Langou J, Kurzak J, et al. A class of parallel tiled linear algebra algorithms for multicore architectures[J]. Parallel Computing, 2009, 35(1): 38-53.

%\bibitem{nn}Anthony M, Bartlett P L. Neural network learning: Theoretical foundations[M]. cambridge university press, 2009.

%\bibitem{onednn} “Optimize AI Applications with Intel® oneAPI Deep Neural Network...” https://software.intel.com/content/www/us/en/develop/tools/oneapi/components/onednn.html

\bibitem{jia}Jia Y. Learning semantic image representations at a large scale[D]. UC Berkeley, 2014.

\bibitem{why} “Why GEMM is at the heart of deep learning” https://petewarden.com/2015/04/20/why-gemm-is-at-the-heart-of-deep-learning/

\bibitem{im2col}Chellapilla K, Puri S, Simard P. High performance convolutional neural networks for document processing[C]//Tenth International Workshop on Frontiers in Handwriting Recognition. Suvisoft, 2006.

\bibitem{numpy}Van Der Walt S, Colbert S C, Varoquaux G. The NumPy array: a structure for efficient numerical computation[J]. Computing in science \& engineering, 2011, 13(2): 22-30.

\bibitem{cache}Frigo M, Strumpen V. The cache complexity of multithreaded cache oblivious algorithms[J]. Theory of Computing Systems, 2009, 45(2): 203-233.

\bibitem{complexity}“Cache-Aware Analysis of Algorithms” https://www.cse.wustl.edu/~angelee/archive/cse341/fall14/handouts/recitation03.pdf

\bibitem{extra1}Cao W, Wang X, Ming Z, et al. A review on neural networks with random weights[J]. Neurocomputing, 2018, 275: 278-287.

\end{thebibliography}
\end{document}